\documentstyle[twocolumn,seceq,epsf]{jpsj}
\def\runtitle{A Unified Theory of Dynamical Mean Field and Spin Fluctuations}
\def\runauthor{Tetsuro {\sc Saso}}

\def\av#1{\langle #1 \rangle}
\def\a{\alpha}
\def\b{\beta}
\def\e{\epsilon}
\def\w{\omega}

\def\chit{\tilde{\chi}}
\def\nt{\tilde{n}}
\def\rt{\tilde{\rho}}
\def\Gt{\tilde{G}}
\def\Pt{\tilde{\Pi}}
\def\Sb{\bar{S}}
\def\s{\sigma}
\def\d{\mbox{d}}
\def\i{\mbox{i}}
\def\ee{\mbox{e}}
\def\Im{\mbox{Im}}
\def\Re{\mbox{Re}}
\def\ii{{\mbox{\scriptsize i}}}

\def\dd{{\mbox{\scriptsize d}}}
\def\AA{{\mbox{\scriptsize A}}}

\def\LL{{\mbox{\scriptsize L}}}
\def\RR{{\mbox{\scriptsize R}}}
\def\PP{{\mbox{P}}}
\def\Ed{\tilde{E}_\dd}

\title
{
A Unified Theory of Dynamical Mean Field and Spin Fluctuations}
\author
{ 
Tetsuro {\sc Saso}\footnote{E-mail: saso@phy.saitama-u.ac.jp}
}
\inst
{
Department of Physics, Faculty of Sciences, Saitama University, Urawa 338-8570
}
\recdate
{
May 25, 1999
}

\abst
{
A new approach is proposed which encompasses the dynamical mean field theory (DMFT) for strongly correlated electron systems and the self-consistent renormalization (SCR) theory of spin fluctuations.
The latter is incorporated into DMFT as a $O(1/d)$ correction, where $d$ is the spatial dimensionality, but in a phenomenological way.
The DMFT is completely recovered when the effect of spin fluctuation is omitted, whereas the properties at the quantum critical point (QCP) is the same as the SCR theory.
An example of calculation is presented for the single band Hubbard model and the non-Fermi liquid behavior is demonstrated at QCP.
}
\kword
{
Dynamical mean field theory, Spin fluctuation
}

\begin{document}
\sloppy
\maketitle

\section{Introduction}
It is one of the most important tasks in condensed matter physics to construct a theory which can describe the electronic states and various physical properties of the strongly correlated electron systems (SCES) including high temperature superconductors, heavy electron systems, organic conductors and so on.
Various numerical methods so far used, {\it e.g.}, numerical renormalization group, density matrix renormalization and quantum Monte Carlo methods, are valid only up to 0, 1 and 2 spatial dimensions, respectively, and extension to three dimensions or to realistic systems with complicated band structure seems difficult.

The dynamical mean field theory\cite{Georges96} (DMFT) is a most powerful method among such theories for describing the effect of strong correlation.
It becomes exact in the limit of large spatial dimension $d \rightarrow \infty$, where the lattice problem is reduced to solving an impurity problem embedded in an effective medium self-consistently.
It can be regarded as a natural extension of the mean field theory for spin systems to the itinerant electron systems and offers a theoretical description of the latter beyond the Hartree-Fock approximation.
Most properties of three dimensional systems, including Kondo insulators,\cite{Saso96,Saso97} are expected to be described well by this method.\cite{Georges96}

DMFT, however, neglects corrections due to finite $d$, which are the fluctuations from the mean field and may become important in the vicinity of the phase transition points.
Several attempts to extend DMFT have been made so far.
Schiller and Ingersent\cite{Schiller95} included $1/d$ correction perturbatively.
Hettler, {\it et al.}\cite{Hettler98} considered a cluster embedded in an effective medium.
These approaches, however, do not seem to have reached a level of practical use.

In the case of the spin fluctuation, the so-called self-consistent renormalization (SCR) theory\cite{Moriya85} is proposed, first for d-electron systems, where the correlation is not very strong.
SCR theory can be microscopically derived,\cite{Moriya73,Kawabata74} but the recent phenomenological construction,\cite{Moriya85} described with only small number of physical parameters, seems much simpler and having wider applicability.
It was later extended to the f-electron systems where the correlation is much stronger.\cite{Moriya95}
This was done also on a phenomenological basis.
Recently, much attention has been paid to the quantum critical point (QCP),\cite{Millis93} at which, {\it e.g.}, antiferromagnetic long range order is destroyed by doping, applying pressure, etc.
It has been proved that SCR can describe physics around QCP successfully.\cite{Kambe96}
On the other hand, since SCR describes a system by only small number of parameters, details of the structure specific to the system may not be taken into account.
Miyake and Narikiyo\cite{Miyake94} extended SCR to include the effect of nesting and the sharp structure of the density of states.
It would be valuable if SCR theory for SCES could be rebuilt on a more microscopic basis.
Purpose of the present paper is to propose such a theory which starts from DMFT and is extended to include SCR, or that which interpolates between DMFT and SCR.

Kuramoto and Miyake\cite{Kuramoto90} proposed a phenomenological theory of heavy electron systems which emphasizes the dual character of the f-electrons and separates the mixture of conduction and f electrons into the localized moment part and the heavy quasi-particle part, which are interacting with each other.
Extending the Fermi liquid theory, they constructed the dynamical susceptibility of the impurity Anderson model, and then proceeded to the duality description of SCES.
Basic assumption of Moriya and Takimoto\cite{Moriya95} in extending SCR to SCES stands on the above theory.

Ohkawa\cite{Ohkawa92,Ohkawa98} discussed an extension of DMFT to include intersite spin fluctuations through $O(1/d)$ correction in detail.
His formulation, however, is rather complicated and needs further approximation or simplification in actual calculations.

Therefore, we propose in this paper a simpler scheme which unifies DMFT and SCR and interpolates between them.
DMFT is treated by the iterative perturbation theory (IPT)\cite{Georges92}, which is simple but valid for wide parameter range of SCES.
The phenomenological construction of SCR, to which its simplicity owes, is kept as much as possible in the present formulation.

The present paper is organized as follows.
In \S 2, the methods of DMFT and IPT are summarized.
In \S 3, a modified IPT scheme, which can be used for electron-hole asymmetric case, is explained.
In \S 4, a unified theory of DMFT and SCR theory is presented.
Numerical calculation for the Hubbard model is presented as an example of the calculation.
Finally, discussion and conclusion are given in \S 5.

\section{Iterative Perturbation Theory}
In the present paper, we investigate the single-band Hubbard model,
\begin{eqnarray}
  H &=& \sum_{ij\s} t_{ij} c_{i\s}^+c_{j\s} + \sum_{i\s} \Ed c_{i\s}^+ c_{i\s}
  + U \sum_i \delta n_{i\uparrow} \delta n_{i\downarrow},\nonumber \\
 \label{eq:ham}
\end{eqnarray}
where $\delta n_{i\s} = n_{i\s}-\av{n_{i\s}}$ and
$\Ed=E_\dd+U\av{n_{-\s}}$ denotes the Hartree-Fock level of d electrons.
$n_{i\s}$ denotes the electron number at site $i$ with spin $\s$ and $\av{\cdots}$ the average.
We do not apply magnetic field, so that $\av{n_{i\s}}$ does not depend on $\s$ nor $i$.
$\av{\cdots}$ will be omitted.
In DMFT, the hopping integral between the nearest neighbor sites is usually scaled as $t_{ij} = t/2\sqrt{d}$.  If we use the simple hyper-cubic lattice, it is well known that the density of states is reduced to the Gaussian form of the width $t$ in the limit $d \rightarrow \infty$.  Furthermore, the lattice problem is reduced to solving the impurity problem embedded in an effective medium self-consistently.
Here, the impurity can be regarded as being placed at the cavity site in the center of the effective medium.
Its unperturbed state is described by the cavity Green's function $\Gt(\e)$, the Green's function at a cavity site (interaction $U$ is artificially removed there), which is related to the local Green's function $G(\e)$ as
\begin{equation}
  \Gt(\e)=(G(\e)^{-1}+\Sigma(\e))^{-1}, \label{eq:Gt}
\end{equation}
and $G(\e)$ is defined by
\begin{equation}
  G(\e)=\frac{1}{N}\sum_k \dfrac{1}{\e-\Ed-\e_k-\Sigma(\e)}. \label{eq:GF}
\end{equation}
$N$ is the number of lattice sites and
$\e_k$ denotes the energy of conduction electrons. $\Sigma(\e)$ denotes the self-energy which describes the effective medium and is self-consistently generated at the cavity site by $U$.

\begin{figure}
\vspace{1cm}
\epsfxsize=4cm
\centerline{\epsfbox{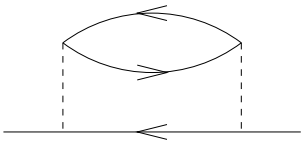}}
\caption{The second-order self-energy.}
\label{fig:self2}
\end{figure}
In IPT, the self-energy $\Sigma(\e)$ is calculated up to the second order of $U$ (Fig.\ref{fig:self2}) as a function of the Matsubara frequency as
\begin{equation}
  \Sigma^{(2)}(\i\e) = U^2 T\sum_\w \Gt(\i\e+\i\w) \Pt(\i\w),
\end{equation}
where $\e=(2n+1)\pi T$, $\w=2n\pi T$ ($n=$integer) and
\begin{equation}
  \Pt(\i\w)=-T\sum_{\e'} \Gt(\i\e') \Gt(\i\e'+\i\w). \label{eq:Pi}
\end{equation}
The self-energy is analytically continuated to the real frequency as
\begin{eqnarray}
  \Sigma^{(2)}(\e) &=& U^2 \int \d \e_1 \int \d \e_2 \int \d \e_3 \ \rt(\e_1) \rt(\e_2) \rt(\e_3) \nonumber \\
  & \hspace{-1cm}\times & \hspace{-5mm}\frac{f(-\e_1)f(-\e_2)f(\e_3)+f(\e_1)f(\e_2)f(-\e_3)}{\e-\e_1-\e_2+\e_3+\i\delta} \label{self}
\end{eqnarray}
where $\rt(\e)=(-1/\pi)\mbox{Im} \Gt(\e+\i\delta)$ is the density of states of the cavity, $\delta$ a small positive number and $f(\e)$ denotes the Fermi function.

This equation is converted into the following form:
\begin{eqnarray}
  \Sigma^{(2)}(\e) &=& -\i U^2\int_0^\infty\!\!\! \d t\ \ee^{\ii\e t} \left[ \b(t) \a(-t) \b(t) \right. \nonumber \\
  & & + \left. \a(t) \b(-t) \a(t) \right],
\end{eqnarray}
by introducing the transformation of $\rt(\e)$:
\begin{equation}
  \left. \begin{array}{l} \a(t) \\ \b(t) \end{array} \right\} = \int_{-\infty}^\infty\!\!\! \d \e\ \ee^{-\ii\e t} \rt(\e) f(\pm\e).  \label{eq:selfb}
\end{equation}
In this paper, we use the semi-circular density of states $\rho(\e)=(2/\pi W)\sqrt{1-(\e/W)^2}$ for the unperturbed electrons.
Then the Green's function, eq.(\ref{eq:GF}), is calculated as
\begin{eqnarray}
G(\e) = \frac{2}{z+\i \sqrt{W^2-z^2}}
\end{eqnarray}
with $z=\e-\Ed-\Sigma(\e)$.
Combining with eq.(\ref{eq:Gt}) we obtain
\begin{equation}
  \Gt(\e)=\frac{1}{\e-\Ed-\frac{W^2}{4}G(\e)}.
\end{equation}
These equations should be calculated self-consistently.

\section{Modified IPT}
IPT reproduces the DMFT calculation via exact diagonalization of the effective impurity in the case of electron-hole symmetry rather well,\cite{Cafferel94} in spite of the use of the second order perturbation.
A reason is firstly that the second-order perturbation for the impurity Anderson model reproduces the atomic limit correctly in the symmetric case.\cite{Yosida70}
Thus it succeeded in describing the Mott transition of the Hubbard model on Bethe lattice at $U/W \approx 3$.\cite{Zhang93}
Recently, it was pointed out that an another reason may be that IPT becomes exact when the width of the conduction band becomes zero.\cite{Lange98}
In the electron-hole asymmetric cases, however, the second-order perturbation does not reproduce the atomic limit, and the positions of the Hubbard bands are wrong.

\begin{figure}
\vspace{1.5cm}
\epsfxsize=8cm
\centerline{\epsfbox{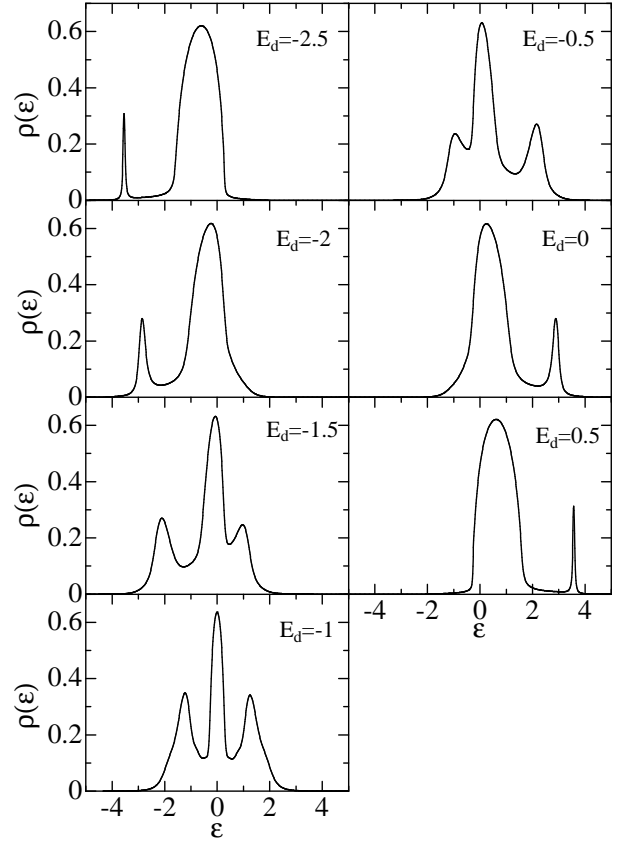}}
\caption{The densities of states calculated by the modified IPT scheme for $W=1$, $U=2$ and various values of $E_\dd$ are displayed. The Fermi energy is set equal to 0.}
\label{fig:rho}
\end{figure}

\begin{figure}
\vspace{1cm}
\epsfxsize=6cm
\centerline{\epsfbox{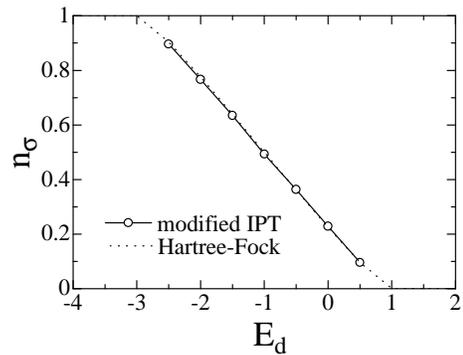}}
\caption{Number of electrons as a function of $E_\dd$ calculated by the Hartree-Fock approximation and by the modified IPT scheme at $T=0$.}
\label{fig:ns}
\end{figure}

To remedy this point, a phenomenological interpolation scheme\cite{LevyYeyati93,Takagi99,Kajueter96,Kajueter96b} was proposed to encompass the perturbative regime and the atomic limit.
In this treatment, the following form is assumed for the modified self-energy:
\begin{equation}
  \Sigma(\e)=\frac{A\Sigma^{(2)}(\e)}{1-B\Sigma^{(2)}(\e)},
\end{equation}
where the parameters $A$ and $B$ are determined so as to reproduce the correct $|\e| \rightarrow \infty$ limit and the atomic limit, respectively, yielding
\begin{equation}
  A=\frac{n_{-\s}(1-n_{-\s})}{\nt_{-\s}(1-\nt_{-\s})},
\end{equation}
\begin{equation}
  B=\frac{1-2n_{-\s}}{U\nt_{-\s}(1-\nt_{-\s})}.
\end{equation}
Here, $\nt_\s$ is the electron number calculated from $\rt(\e)$ instead of $\rho(\e)=-(1/\pi)\Im G(\e+\i\delta)$.
Then, the Green's function $G(k,\e)=(\e-\Ed-\e_k-\Sigma(\e))^{-1}$ reduces to that of the atomic limit
\begin{equation}
  G_a(\e)=\frac{1-n_{-\s}}{\e-E_\dd}+\frac{n_{-\s}}{\e-E_\dd-U},
\end{equation}
when $U\gg W$,
but does not satisfies the Luttinger sum rule because of the phenomenological modification of the self-energy.  This point can be remedied by tuning $\Ed$ as the effective level so as to satisfy the Luttinger sum rule.\cite{Kajueter96,Kajueter96b}
We, however, adopt a more simplified scheme, in which the sum rule is satisfied only approximately.
Nevertheless, physically reasonable behavior is obtained for a wide range of parameters without tedious determination of the effective $\Ed$.
This is achieved by simply subtracting $\Re \Sigma(0)$\cite{Yamada79} at $T=0$ from $\Sigma(\e)$ at each temperature.
We show the density of states for various values of $E_\dd$ with $W=1$ and $U=2$ in Fig. \ref{fig:rho}.
The Fermi energy is set equal to 0.
It is seen that almost correct shapes of the spectra are obtained for all cases.
The Luttinger sum rule is stated as $n_\s^\LL=n_\s$,
if we define the electron number $n_\s^\LL$ by
\begin{equation}
  n_\s^\LL=\sum_k \theta(\mu-\Ed-\Sigma(\mu)-\e_k),
\end{equation}
where $\mu$ denotes the chemical potential.
In Fig.\ref{fig:ns}, $n_\s$ in the modified IPT and in the Hartree-Fock approximation (HFA) are plotted as functions of $E_\dd$.
Since $\Sigma(\mu)=0$ in our treatment, $n_\s^\LL$ is approximately equal to $n_\s$ in HFA, whereas $n_\s \simeq n_\s$(HFA) in Fig.\ref{fig:ns}.
Therefore, the Luttinger sum rule is approximately fulfilled in the present treatment.

\section{Inclusion of Spin Fluctuations}
In this section, we first discuss the dynamical susceptibility of SCES.
In extending SCR theory to the case of SCES,
Moriya and Takimoto\cite{Moriya95} assumed the following form for the dynamical susceptibility, motivated by the duality picture of Kuramoto and Miyake:\cite{Kuramoto90}
\begin{equation}
  \chi(Q+q,\w)=[\chi_\LL(\w)^{-1}-J_Q(T)+Aq^2]^{-1}, \label{eq:chiqw}
\end{equation}
where $\chi_\LL(\w)$ denotes the local susceptibility of each magnetic ion and $Q$ an ordering vector.
In SCR, $\chi_\LL(\w)^{-1}$ is expanded up to the linear term of $\w$, yielding
\begin{equation}
\chi_\LL^{\mbox{\scriptsize SCR}}(\w)=\frac{\chi_L}{1-\i\w/\Gamma_\LL}, 
\end{equation}
below certain cutoff frequency $\w_c$.
Here, $\Gamma_\LL$ is of the order of the Kondo temperature of the magnetic ion.

\begin{figure}
\vspace{1cm}
\epsfxsize=8cm
\centerline{\epsfbox{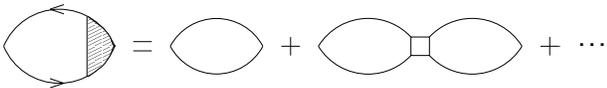}}
\caption{Feynman diagram for $\chi_\LL$ is show where the bubble denotes $\Pi$ and the square denotes $\Gamma$.}
\label{fig:chi}
\end{figure}

Kuramoto and Miyake\cite{Kuramoto90} derived an approximate form of the dynamical susceptibility for the impurity Anderson model on the basis of the Fermi liquid theory. 
We calculate it here in a similar but slightly different manner as
\begin{equation}
  \chi_\LL(\w)=\frac{\Pi(\w)}{1-\Gamma(\w)\Pi(\w)}, \label{eq:chi}
\end{equation}
where $\Pi(\w)$ is defined in the same way as $\Pt(\w)$ in eq.(\ref{eq:Pi}) but with $G(\e)$ instead of $\Gt(\w)$.
The Feynman diagram is shown in Fig.\ref{fig:chi}.
$\Gamma(\w)$ is the renormalized vertex function, for which we adopt an approximate form,
\begin{equation}
  \Gamma(\w)=\frac{U}{1+U\Pi(\w)},
\end{equation}
which corresponds to taking fan type vertex corrections into account.
This form of $\Gamma(\w)$ has a useful property.
Since $\Gamma(\w) \rightarrow \Pi(\w)^{-1}$ in $U \rightarrow \infty$, the denominator of $\chi_\LL(\w)$ in eq.(\ref{eq:chi}) does not vanish as far as $U < \infty$ and the magnetic instability is avoided.
This is a necessary property when one wants to calculate $\chi(\w)$ of the single Anderson impurity for large $U$.
Eq.(\ref{eq:chi}) can be rewritten as
\begin{equation}
  \chi_\LL(\w)=\Pi(\w)[1+U\Pi(\w)],
\end{equation}
which reproduces precisely the perturbation series expansion of $\chi(\w=0)$ up to the first order of $U$ for the symmetric and asymmetric cases of the Anderson model.\cite{Yamada79}

In contrast to SCR theory, the present theory retains full dynamical structure of $\chi_\LL(\w)$.
But because of the approximate vertex correction, $\chi_\LL(\w)$ does not satisfy the sum rule,
\begin{equation}
  S_\LL^2=\frac{3}{\pi}\int_0^\infty \!\!\! \d \w [1+2n(\w)] \Im \chi_\LL(\w) = S(S+1),
\end{equation}
where $S=1/2$.
In order to fulfill this sum rule, we modify $\chi_\LL(\w)$ as follows,
\begin{equation}
  \chi'_\LL(\w)=[\chi_\LL(\w)^{-1}-\i C\w]^{-1},
\end{equation}
with the parameter $C(T)$ to be adjusted at each temperature.
This may correspond to modifying $\Gamma_\LL$ to an effective value to satisfy the sum rule.
Here and henceforth, we omit the prime on $\chi_\LL$.

\begin{figure}
\vspace{1cm}
\epsfxsize=6cm
\centerline{\epsfbox{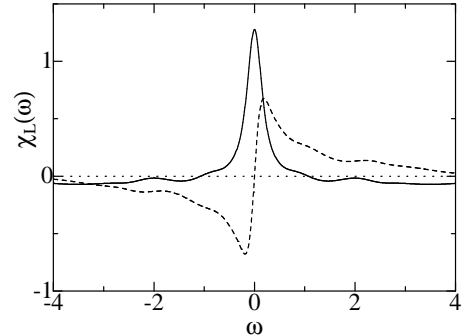}}
\caption{The real (solid line) and imaginary (broken line) parts of the local susceptibility $\chi_\LL(\w)$ are plotted for $U=2$, $\Ed=-0.5$ and $T=0$.}
\label{fig:chiLw}
\end{figure}

\begin{figure}
\vspace{1cm}
\epsfxsize=6cm
\centerline{\epsfbox{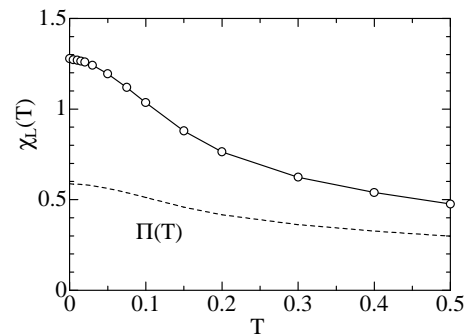}}
\caption{Temperature dependence of the local susceptibility $\chi_\LL(\w=0,T)$ is plotted.  $\Pi(\w=0,T)$ is also shown by the broken line for comparison.}
\label{fig:chiLT}
\end{figure}

We show numerical results for $\chi_\LL(\w)$ at $T=0$ and $\chi_\LL(\w=0,T)$ in Figs.\ref{fig:chiLw} and \ref{fig:chiLT}, respectively.  
We have used the IPT calculation for $E_\dd=-0.5$ in the preceeding section.
For comparison, we also show $\Pi(\w=0,T)$ in Fig.\ref{fig:chiLT}.
It is seen that $\chi_\LL(T)$ is enhanced over $\Pi(T)$ due to the vertex correction.

In DMFT, the dynamical susceptibility $\chi(q,\w)$ of the lattice can be obtained once the effective impurity problem was solved.\cite{Georges96}
The procedure, however, needs rather tedious calculations.
Ohkawa\cite{Ohkawa98} and Miyake and Narikiyo\cite{Miyake94} discussed a general form of $\chi(q,\w)$ and derived the following equation
\begin{equation}
  \chi(q,\w)=[\chi_\LL(\w)^{-1}-J(q,\w)]^{-1} \label{eq:chiqwO}
\end{equation}
where $J(q,\w) \simeq U^2 \Delta\chi_0(q,\w) + \lambda(q,\w)$ in the strong correlation limit.
$\Delta\chi_0(q,\w)$ denotes the intersite part of the dynamical susceptibility of the lattice,
\begin{equation}
  \chi_0(q,\i\w)=-T\sum_{k,\e'} G(k,\i\e') G(k+q,\i\e'+\i\w),
\end{equation}
\begin{equation}
  \Delta\chi_0(q,\i\w) = \chi_0(q,\i\w) - \frac{1}{N}\sum_q \chi_0(q,\i\w),
\end{equation}
and $\lambda(q,\w)$ contribution from the mode-mode coupling of spin fluctuations.
$J(q,\w)$ includes both the mean field contribution to the exchange interaction, which is $O(1/d^0)$, and the fluctuations from the mean field, which is $O(1/d)$.
Here we assume that $J(q,\w)$ can be expanded around a certain ordering vector $Q$ as $J(Q+q,\w)=J_Q(T)-Aq^2 + \cdots$ and $\w$-dependence can be neglected.
Then we obtain $\chi(Q+q,\w)$ in the same form as eq.(\ref{eq:chiqw}).
We can determine $J_Q(T)$ by the condition that the spin-fluctuation amplitude
\begin{equation}
  \Sb_\LL^2=\frac{3}{\pi}\int_0^\infty \d \w [1+2n(\w)] \Im \chi_Q(\w) \label{eq:SL}
\end{equation}
stays constant: $\Sb_\LL^2(T_N)=\Sb_\LL^2(T)$, where $T_N$ is the ordering temperature.
Both quantum and thermal fluctuations are included in $\Sb_\LL^2(T)$.
When there is no long range order, we set $T_N=0$ in the above condition.
In eq.(\ref{eq:SL}), $\chi_Q(\w)$ is defined by
\begin{eqnarray}
  \chi_Q(\w)&=&\frac{1}{N}\sum_q \chi(Q+q,\w) \nonumber \\
            &=&\frac{3}{T_\AA}[1-\sqrt{y_Q(\w)}\tan^{-1}\frac{1}{\sqrt{y_Q(\w)}}],
\end{eqnarray}
where $y_Q(\w)=[\chi_\LL(\w)^{-1}-J_Q(T)]/T_A$, $T_A=Aq_B^2$ characterizes the stiffness of the spin fluctuation and $q_B$ is the wave vector at the Brillouin zone boundary.
In the actual calculation we take $q_B^3=6\pi^2N/V$, where $V$ is the system volume.

\begin{figure}
\vspace{1cm}
\epsfxsize=6cm
\centerline{\epsfbox{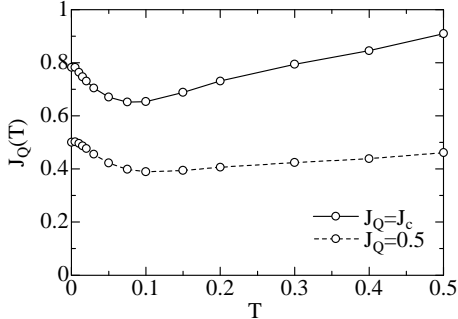}}
\caption{The temperature dependence of $J_Q(T>0)$ is plotted for the cases $J_Q(T=0)=0.5$ and $J_c$.}
\label{fig:JQT}
\end{figure}

In Fig.\ref{fig:JQT}, we show the temperature dependence of the parameter $J_Q(T)$ when $J_Q \equiv J_Q(T=0)=0.5$ and 0.7827$\equiv J_c$, which is a critical value and is equal to $\chi_\LL(\w=0,T=0)^{-1}$.
$J_Q(T>0)$ is determined by the condition $\Sb_\LL^2(0)=\Sb_\LL^2(T)$.

\begin{figure}
\vspace{1cm}
\epsfxsize=6cm
\centerline{\epsfbox{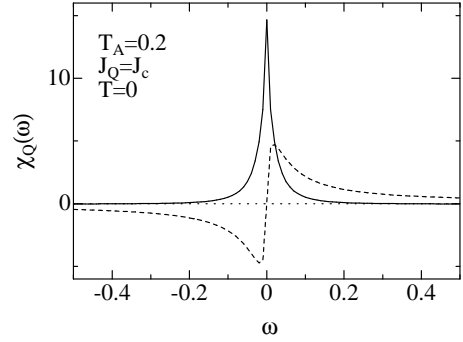}}
\caption{The real (solid line) and imaginary (broken line) parts of the dynamical susceptibility $\chi_Q(\w)$ at $T=0$ are plotted.}
\label{fig:chiQw}
\end{figure}

\begin{figure}
\vspace{1cm}
\epsfxsize=6cm
\centerline{\epsfbox{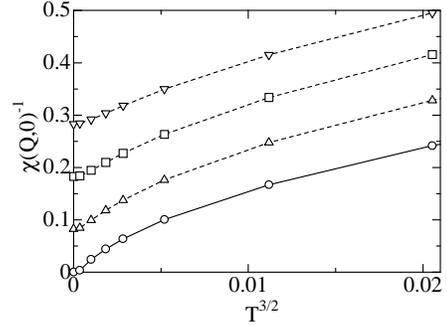}}
\caption{The susceptibility $\chi_Q(0)$ is plotted as a function of $T$ for $J_Q=$0.5, 0.6, 0.7 and $J_c$ from upper to lower curves.}
\label{fig:chiQT}
\end{figure}

The frequency dependence of $\chi_Q(\w)$ and the temperature dependence of $\chi(Q,0)$ are plotted in Figs.\ref{fig:chiQw} and \ref{fig:chiQT} for $J_Q=$0.5, 0.6, 0.7 and $J_c$.
It is seen that the low energy structure of $\chi_Q(\w)$ becomes singular and much narrower than $\chi_\LL(\w)$, and $\chi(Q,0)$ becomes proportional to $T^{3/2}$ at QCP, as expected.\cite{Moriya95}

\begin{figure}
\vspace{1cm}
\epsfxsize=8cm
\centerline{\epsfbox{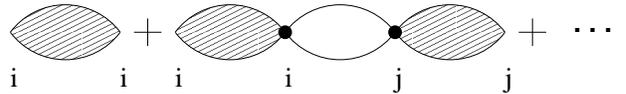}}
\caption{Feynman diagram for $\chi_{ij}$ is shown, where the hatched bubble denotes $\chi_\LL$ and the white bubble $\Delta\chi^0_{ij}$. The filled circles denote $U$.}
\label{fig:chi-ij}
\end{figure}

Once these calculations are done, we want to recalculate the self-energy.
It must be done in a manner consistent with the IPT calculation.
The dynamical susceptibility $\chi_{ij}(\w)$ in the site representation may be expanded as
\begin{eqnarray}
  \chi_{ij} &=& \chi^*_\LL \delta_{ij} + U^2 \chi^*_\LL \Delta\chi^0_{ij} \chi^*_\LL \cr
 & + & U^4 {\sum_k}' \chi^*_\LL \Delta\chi^0_{ik} \chi^*_\LL \Delta\chi^0_{kj} \chi^*_\LL  + \cdots -U^{-1},
\end{eqnarray}
(Fig.\ref{fig:chi-ij}), where $\chi^0_{ij}$ denotes the irreducible susceptibility, $\chi^*_\LL=1/U+\chi_\LL$ and the contributions from the successive identical site are collected into $\chi_\LL \equiv \chi^0_{ii}/(1-U\chi^0_{ii})$, whereas the successive identical sites in the intermediate paths are excluded in the summation as indicated by the prime since $\Delta \chi^0_{ii}=0$.
The above equation can be solved after Fourier transformation into
\begin{equation}
  \chi(q,\w) = \frac{\chi^*_\LL(\w)}{1-U^2\Delta\chi_0(q,\w)\chi^*_\LL(\w)},
\end{equation}
which can be rewritten in the same form as eq.(\ref{eq:chiqwO}), if we define $J(q,\w)\equiv U^2\Delta\chi_0(q,\w)$ and note that $\chi^*_\LL \simeq \chi_\LL$ in SCES.\cite{Ohkawa98}

It may seem that one can calculate the self-energy of electrons as scattered by this spin fluctuation, but
the calculation of the self-energy at the cavity site needs further modification.
Namely, we divide eq.(\ref{eq:chiqwO}) into the form
\begin{equation}
  \chi(q,\w) = \chi_\LL(\w) + \frac{J(q,\w)\chi_\LL(\w)^2}{1-J(q,\w)\chi_\LL(\w)},
\end{equation}
and replace $\chi_\LL(\w)$ in the first term and in the numerator of the second term with $\Pt(\w)$.
Strictly speaking, the paths which go through the cavity site in the intermediate steps must be subtracted,\cite{Smith97} but we neglect that effect in the present study.
Then we obtain
\begin{equation}
  \Sigma(\i\e)=U^2 T\sum_\w \Gt(\i\e+\i\w) \chit_Q(\i\w), \label{eq:self}
\end{equation}
$$  \chit_Q(\i\w) = \Pt(\i\w) + \frac{1}{N} \sum_q \frac{J(Q+q,\i\w)\Pt^2(\i\w)}{1-J(Q+q,\i\w)\chi_\LL(\i\w)} $$
\begin{equation}
  = \Pt(\i\w) + [\chi_Q(\i\w)-\chi_\LL(\i\w)] \left(\frac{\Pt(\i\w)}{\chi_\LL(\i\w)}\right)^2, \label{eq:chitQ}
\end{equation}
where we have replaced $J(q,\w)$ with $J_Q(T)-Aq^2$.
Note that the second term has the same criticality as $\chi_Q(\w)$.
If we set $J(Q+q,\i\w)=0$ in this form, the self-energy eq.(\ref{eq:self}) recovers the IPT result.

\begin{figure}
\vspace{1cm}
\epsfxsize=6cm
\centerline{\epsfbox{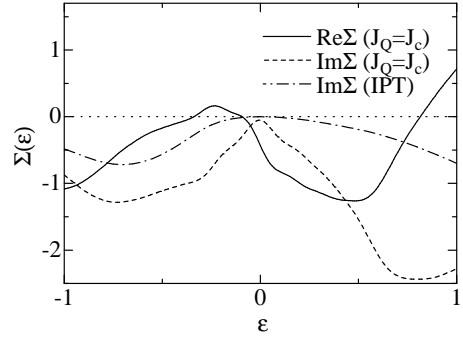}}
\caption{The real (solid line) and imaginary (broken line) parts of the self-energy at $T=0$ is shown for $J_Q=J_c$. The imaginary part of the self-energy in IPT is also shown by the dash-dotted line for comparison.}
\label{fig:self}
\end{figure}

For numerical calculation, the following expression is convenient (see Appendix):
\begin{equation}
  \Sigma(\e) = -\i U^2\int_0^\infty\!\!\! \d t\ \ee^{\ii\e t} \left[ \b(t) \gamma(t) + \a(t) \gamma(-t) \right], \label{eq:Sigma}
\end{equation}
where
\begin{equation}
  \gamma(t) = \frac{1}{\pi}\int_{-\infty}^\infty\!\!\! \d \w\ \ee^{\ii\w t} n(\w) \Im \chit_Q(\w+\i\delta).
\end{equation}
The low energy regions of real and imaginary parts of $\Sigma(\e)$ are shown in Fig.\ref{fig:self}, and the imaginary part is compared with that by IPT.
It is seen that $\Im \Sigma(\e)$ at $\e\sim 0$ becomes singular in contrast to the Fermi liquid behavior $\Im \Sigma(\e) \propto \e^2$ in IPT.
Theoretical analysis leads to $\Im \Sigma(\e) \propto \e^{3/2}$ at QCP and $T=0$.

\begin{figure}
\vspace{1cm}
\epsfxsize=6cm
\centerline{\epsfbox{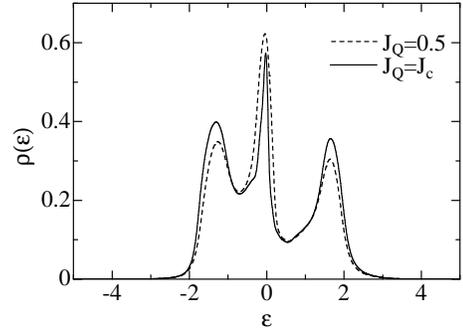}}
\caption{The density of states at $T=0$ after inclusion of the spin fluctuation.  The broken line is for $J_Q=0.5$ and solid line for $J_Q=J_c$}
\label{fig:rhoL}
\end{figure}

Using the above self-energy, we calculate the density of states again, which is shown in Fig.\ref{fig:rhoL}.
It is seen that the peak at $E_F$ becomes thinner and singular due to the strong renormalization by the spin fluctuation.

\begin{figure}
\vspace{1cm}
\epsfxsize=6cm
\centerline{\epsfbox{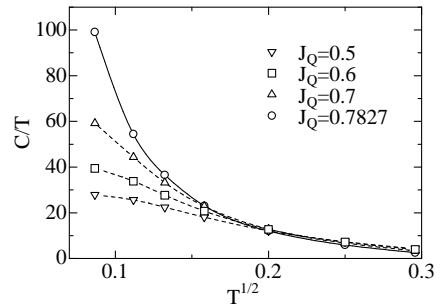}}
\caption{Temperature dependence of $C(T)/T$ is plotted for $J_Q=$0.5, 0.6, 0.7 and $J_c$.}
\label{fig:c-t}
\end{figure}

Finally, the total energy is calculated from\cite{Fetter71}
\begin{equation}
 E = 2\int_{-\infty}^\infty \!\!\!\d \e f(\e) \left(-\frac{1}{\pi}\right) \Im [\{\e-\frac{1}{2}(Un_{-\s}+\Sigma(\e))\} G(\e)].
\end{equation}
The specific heat is obtained by numerical derivative of $E$ as $C=\partial E/\partial T$, which is plotted in Fig.\ref{fig:c-t}.
We see that $C(T)/T \propto -\sqrt{T}$ at QCP as in SCR.\cite{Moriya95}

\begin{figure}
\vspace{1cm}
\epsfxsize=6cm
\centerline{\epsfbox{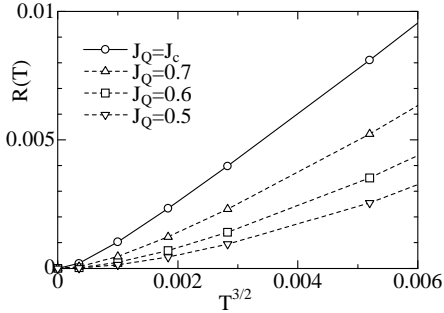}}
\caption{Temperature dependence of the resistivity is plotted for $J_Q=$0.5, 0.6, 0.7 and $J_c$.}
\label{fig:r-t}
\end{figure}

The electrical resistivity in the dimensionless form is calculated by the formula,\cite{Moriya95}
\begin{equation}
  R(T)= \int_0^\infty \!\!\!\d \w \ n(\w) [1+n(\w)] \Im \chi_Q(\w).
\end{equation}
Since this formula is an approximate one, we do not take care whether we should use $\chit_Q(\w)$ in stead of $\chi_Q(\w)$.
Furthermore, the behaviors of $\chi_Q(\w)$ and $\chit_Q(\w)$ are the same at the critical point.
The numerical result is shown in Fig.\ref{fig:r-t}, which seems consistent with the theoretical analysis $R(T) \propto T^{3/2}$ within numerical accuracy of the calculation.\cite{Moriya95}

\section{Discussion and Conclusion}
In this paper, we have proposed a new scheme which incorporates the dynamical mean field theory and the self-consistent renormalization theory of spin fluctuations.
Starting from the IPT calculation (eqs.(2.7-10) and (3.1-3)) for DMFT, we calculated the renormalized local dynamical susceptibility $\chi_\LL(\w)$ approximately (eqs.(4.5-7)), and then constructed the dynamical susceptibility $\chi(q,\w)$ of the whole system in the same form as Ohkawa\cite{Ohkawa92,Ohkawa98} and Moriya and Takimoto.\cite{Moriya95}
Considering the contribution of the mode-mode coupling in a phenomenological manner, we determined its temperature dependence by the sum rule in parallel with the SCR theory (eqs.(4.11) and (4.12)).
Both quantum and thermal spin fluctuations are taken into account.
Since the theory is based on IPT, it is applicable to the strongly correlated systems.
The self-energy of electrons due to the spin fluctuations is calculated by using the above $\chi(q,\w)$ with a correction at the cavity site in the mean field (eqs.(4.17-19)), which ensures a consistency with IPT.
The present theory encompasses the IPT calculation for DMFT and the SCR theory.
If we omit $J(q,\w)$, we recover the IPT results, whereas the critical behavior at QCP is completely the same as SCR.
Therefore, structures in the single-particle spectra specific to each material can be taken into account through the IPT calculation.
Application to the Kondo insulators is also possible.
Combination with band calculations will be an important subject in the future studies.
When the nesting of the Fermi surface is strong, one has to take account of the frequency dependence of $J(q,\w)$.\cite{Miyake94}.

The present treatment still includes phenomenological constructions in incorporating the $1/d$ correction to DMFT.
Furthermore, the self-energy is not calculated self-consistently, together with the effect of spin fluctuations.
Namely, the $1/d$ correction was taken into account only in a perturbative manner.
Recently, several attempts are done to systematically take account of the $1/d$ corrections to DMFT.\cite{Schiller95,Hettler98,Smith97}
Among them, Smith and Si\cite{Smith97} proposed an extended mean field theory which includes the spin-spin coupling in a non-trivial way in addition to the effective impurity action.
Another important development in the effort to investigate the effect of fluctuations may be the two-particle self-consistent (TPSC) theory,\cite{Vilk94,Dare96} which leads to a similar theory to SCR but can take account of both the charge and spin fluctuations.
Combination of the present formulation with such theories may open a way to an extension of DMFT in a more microscopic fashion.

\section*{Acknowledgements}
The author thanks Mr. I. Horikoshi and Mr. T. Kitajima for their contributions at the early stage of the present study.
This work is supported by Grant-in-Aid for Scientific Research No.11640367 
from the Ministry of Education, Science, Sports and Culture.
The numerical computation was partly done using FACOM VPP500 in the Supercomputer Center,
Institute for Solid State Physics, University of Tokyo.

\appendix
\section{Derivation of eq.(\ref{eq:Sigma})}
By analytic continuation, eq.(\ref{eq:self}) is written as
\begin{eqnarray}
  \Sigma(\e) = U^2\int_{-\infty}^\infty\frac{\d \w}{\pi} [\PP n(\w)\Gt^\RR(\e+\w)\Im \chit_Q^\RR(\w) \nonumber \\
  -f(\e+\w)\Im\Gt^\RR(\e+\w)\chit_Q^\AA(\w)].
\end{eqnarray}
Using the spectral representation for $\Gt(\e)$ and $\chit_Q(\w)$, we obtain
\begin{equation}
  \Sigma(\e)=U^2\int\d\w\int\d\nu \frac{n(\w)+f(\nu)}{\e+\w-\nu+\i\delta}\rt(\nu)\xi(\w),
\end{equation}
where $\xi(\w)\equiv(1/\pi)\Im\chit\RR(\w)$.
Introducing
\begin{equation}
  \frac{1}{\e+\w-\nu+\i\delta} = -\i \int_0^{\infty} \d t\ e^{\ii(\e+\w-\nu+\ii\delta)t}
\end{equation}
and
\begin{eqnarray}
  \gamma(t)  &=& \int_{-\infty}^\infty \d\w\ e^{\ii\w t} \xi(\w)n(\w), \\
  \gamma_1(t) &=& \int_{-\infty}^\infty \d\w\ e^{\ii\w t} \xi(\w)[n(\w)+1],
\end{eqnarray}
we obtain
\begin{eqnarray}
  \Sigma(\e) = -\i U^2 \int_0^\infty \d t\ e^{\ii \e t} [(\a(t)+\b(t))\gamma(t) \nonumber \\
  +\a(t)(\gamma_1(t)-\gamma(t))].
\end{eqnarray}
But since we can easily prove $\gamma_1(t)=\gamma(-t)$, we finally obtain eq.(\ref{eq:Sigma}).

\end{document}